\pdfoutput=1
\documentclass[showpacs,prl,preprint]{revtex4}
\usepackage{amsmath}
\usepackage{amssymb}
\usepackage{color}
\usepackage{graphics}
\usepackage{graphicx}
\usepackage{dcolumn}
\usepackage{pdfsync}

\def\q{\quad}

\begin{document}

\title {Fields radiated by a nanoemitter in a graphene sheet}

\author{ A.~Yu.~Nikitin$^{1,2}$}
\email{alexeynik@rambler.ru}
\author{ F. Guinea$^3$}
\author{ F.~J.~Garc\'{i}a-Vidal$^4$}
\author{ L.~Mart\'{i}n-Moreno$^1$}
\email{lmm@unizar.es}
 \affiliation{$^1$ Instituto de Ciencia de Materiales de Arag\'{o}n and Departamento de F\'{i}sica de la Materia Condensada,
CSIC-Universidad de Zaragoza, E-50009, Zaragoza, Spain \\
$^2$A.Ya. Usikov Institute for Radiophysics and Electronics, Ukrainian Academy of
Sciences, 12 Acad. Proskura Str., 61085 Kharkov, Ukraine\\
$^3$ Instituto de Ciencia de Materiales de Madrid, CSIC, Cantoblanco,
E-28049 Madrid, Spain\\
$^4$ Departamento de F\'{i}sica Te\'{o}rica de la Materia Condensada, Universidad Aut\'{o}noma de Madrid, E-28049
Madrid, Spain}

\begin{abstract}
The extraordinary properties of graphene make it a very promising material for use in optoelectronics. However,
this is still a nascent field, where some basic properties of the electromagnetic field in graphene must be explored.
Here we report on the fields radiated by a nanoemitter lying on a graphene sheet. Our results show that this field presents a rich dependence on both frequency, distance to the source and dipole orientation. This behavior is attributed to distinct peculiarities on the density of electromagnetic states in the graphene sheet and the interaction between them. The field is mainly composed of an core region of high-intensity electromagnetic field, dominated by surface plasmons, and an outer region where the field is practically the same it would be for an emitter in vacuum. Within the core region, the intensity of the electric field is several orders of magnitude larger than what it would be in vacuum. Importantly, the size of this core region can be controlled thorough external gates, which opens up many interesting applications in, for instance, surface optics and spectroscopy. Additionally, the large coupling between nanoemitters and surface plasmons makes graphene sheets a propitious stage for quantum-optics, in which the interaction between quantum objects could be externally tailored at will.

\end{abstract}

 \maketitle


{\it Introduction.-}
Graphene has attracted recently a great deal of attention due to its amazing electronic properties \cite{Novoselov05,Review09}.  On top of fundamental issues, such as being the thinnest possible two-dimensional electron gas (2DEG) and possessing charged massless quasiparticles, graphene presents very interesting material properties for its use in electronics. Specially noteworthy are both the possibility of controlling the chemical potential, $\mu$ (and thus the conductivity $\sigma$) through gate voltages and its high mobility at room temperature. Also the optical properties of graphene are exceptional \cite{Geim09,Bonaccorso10}, as they allow the visualization of a material that is just one atom-thick, and even differentiate regions with different number of atomic layers \cite{Casiraghi07,Blake07}. Remarkably, even one-atom thick graphene can bind electromagnetic (EM) modes, despite being almost transparent. These surface EM modes are surface plasmon polaritons (SPP) with a transverse-magnetic (TM) polarization
\cite{Shung86, Campagnoli89,Vafek06,Hansonw08} when  $\mathrm{Im}{\sigma}>0$, which occurs below a critical frequency $\omega_0$ that depends on $\mu$, thus being externally tunable.
For frequencies above $\omega_0$, $\mathrm{Im}{\sigma}<0$ and graphene supports transverse electric (TE) bound EM modes \cite{MikhalkovPRL07}, which are reminiscent of the guided modes in a dielectric film. The existence of SPPs in graphene, with their similarities to those supported by metal surfaces, brings to the fore the possibility of using graphene for many of the functionalities sought within the field of Plasmonics. For instance, graphene could be used to built terahertz (THz) switches \cite{Bludov10}, and a graphene sheet with a spatially non-uniform conductivity is a promising platform for both THz flatland metamaterials and transformation optics \cite{Engheta11}.

\begin{figure}[thb!]
\includegraphics[width=10cm]{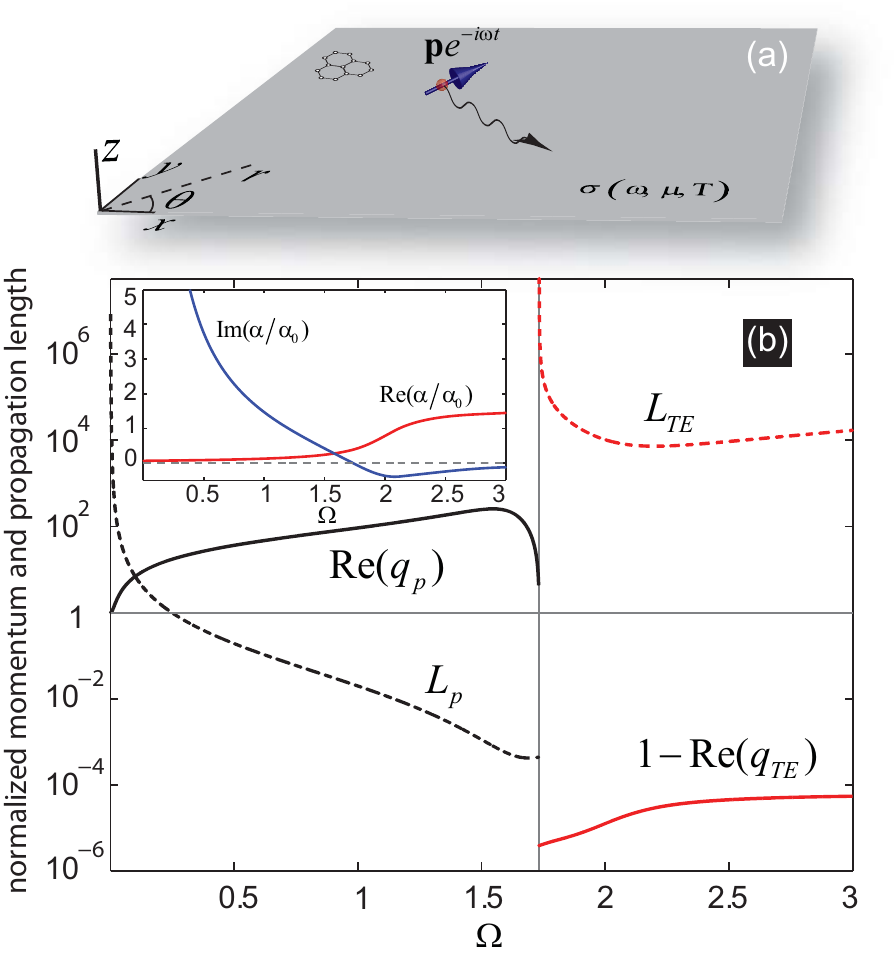}
\caption{
(a) The geometry of the studied system.
(b) Dispersion relation $q(\Omega)$ (continuous curves)
and propagation length (discontinuous curves) of SPPs and TE bound modes, where $q$ is the wavevector in units of $\omega/c$ and $\Omega = \hbar \omega/\mu$. These magnitudes are obtained for the
case in which $\mu=0.2 eV$ (48 THz) and $T=300K$.
The inset shows $\alpha=2\pi\sigma/c$ in units of the fine structure constant $\alpha_0\simeq1/137$.} \label{Figconductivity}
\end{figure}

For all these optical functionalities, an essential requirement is the {\it efficient excitation} of the surface EM modes. In this work we analyze how these bound modes are excited by an emitter of deep subwavelength dimensions (quantum dots, molecules, dielectric or metallic protuberances, etc.) in graphene. Additionally, we study the EM field pattern at the graphene layer, which is an essential ingredient for the understanding of the EM properties of nanostructures placed on a graphene sheet and the effective EM interaction between them. Our results show that the field patterns are drastically different for different frequency ranges and orientation of the dipole. While a nanoemitter hardly couples to bound TE modes, SPPs in graphene can be efficiently excited in a frequency range that depends on the chemical potential of the 2DEG, which allows the control of their excitation by external gates. This suggests graphene as an excellent candidate for supporting adressable long-distance entanglement between qubits  \cite{alejandro11}.

\par


{\it Description of the system under study.-}
We consider a point emitter (placed at the origin, with dipole moment $\mathbf{p}$) lying on a free-standing graphene sheet (covering the plane $z=0$). We will concentrate on the study of the electric field $\mathbf{E}$ {\it at} the graphene sheet, which is the relevant quantity concerning the coupling of subwavelength objects placed on graphene.
Graphene is represented by its in-plane complex conductivity $\sigma$, that is a function of both frequency, $\omega$, and material properties, such as chemical potential $\mu$ and temperature $T$.
In this work, we consider $T=300$K and $\mu=0.2$eV (48THz), which are typical experimental values. The conductivity $\sigma$ is taken from calculations based on the random-phase-approximation
\cite{Wunsch06,Hwang07,Falkovsky08}, and is represented in the inset to Fig. 1b. For this set of parameters the imaginary part of the conductivity vanishes at $\hbar \omega_0= 0.32 eV$. Nevertheless, we have checked that all results presented in this paper are qualitatively valid for other values of $\mu$ (provided that $\omega$ is scaled in the same manner) and $T$.
We have also checked that our results hold true under the presence of a dielectric substrate, if the substrate does not support optical resonances in the frequency range of interest. The interaction of EM modes in graphene with those of a polar substrate is beyond the scope of this paper and is left for future work.

\par

\begin{figure}[thb!]
\includegraphics[width=10cm]{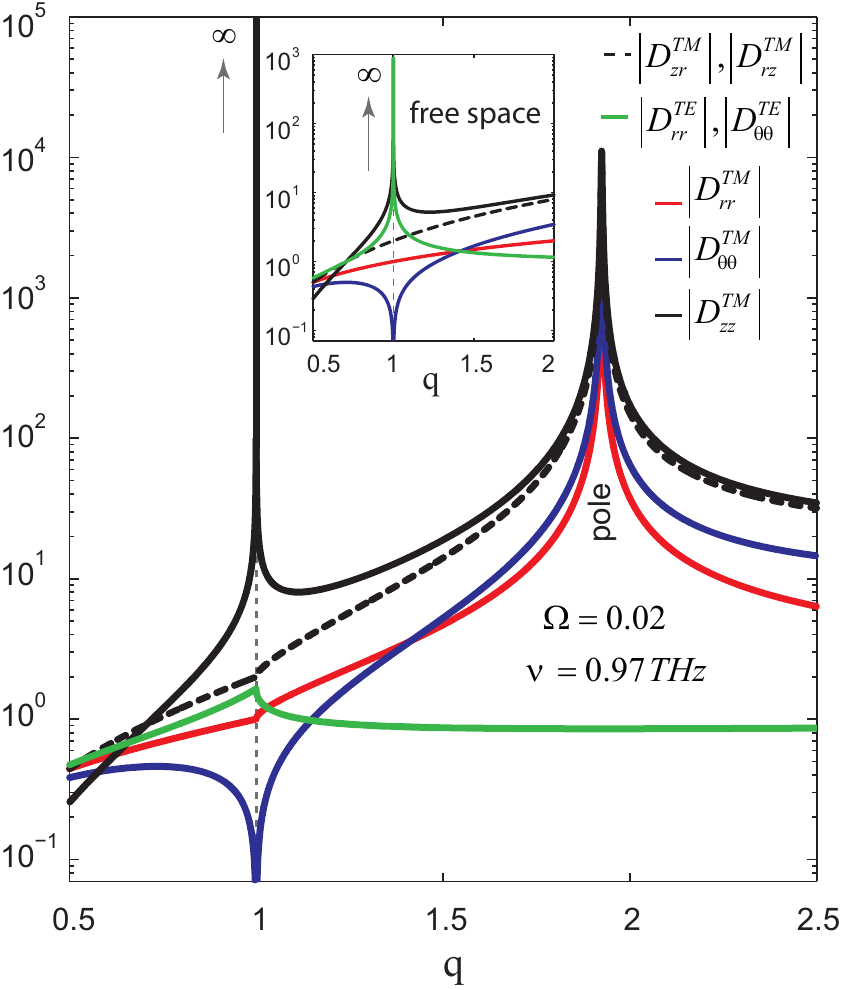}\\
\caption{The different components of the angular spectrum Dyadic $\hat{D}$ as a function of the normalized in-plane wavevector $q= c k/ \omega$. The main figure corresponds to the graphene sheet (for the case $T=300K$ and $\mu=0.2eV$) while the inset is for vacuum. In both cases $\Omega=0.02$, which corresponds to $\nu \approx 1$THz ($\lambda=0.3$mm) for the $\mu$ considered. }\label{Figdensity}
\end{figure}

{\it Mathematical formulation.-}
The electric field radiated by the dipole can be written as  $\mathbf{E}(\mathbf{r}) = \hat{G} (\mathbf{r}) \, \mathbf{p}$, where $\hat{G}(\mathbf{r})$ is the Green's dyadic of the problem. We express distances $R$ in dimensionless units as $r=k_\omega R = 2 \pi R/ \lambda$, being $k_\omega= \omega/c$, $c$ the speed of light and $\lambda$ the wavelength of the EM field. Similarly, we define dimensionless quantities for wavevectors (through $q=k/k_\omega$), conductivity ($\alpha=2 \pi \sigma/c$) and frequency ($\Omega=\hbar \omega / \mu$). Due to the symmetry of the problem, it is convenient to work in cylindrical coordinates, $\mathbf{r}=(r,\theta)$ (see Fig. 1).
Following standard techniques in electromagnetism (see for instance \cite{Novotny}), $\hat{G}$ can be represented in terms of Sommerfeld integrals:

\begin{equation}\label{e3}
\hat{G}_{ij}= \frac{ik_\omega}{8\pi} \sum_{\tau=TE,TM} \int_0^\infty \, dq  \,
\hat{D}_{ij}^\tau(q) \, \hat{J}_{ij}^\tau(qr)
\end{equation}

\noindent where $i,j$ are $r, \theta$ or $z$, $\hat{J}_{ij}^\tau(qr)$ are combinations of Bessel functions and $D_{ij}^\tau$ are the components of the angular spectrum dyadic (see the Supplementary Material for the derivation and explicit expressions).
The non-zero components of $D_{ij}^\tau$ are represented in Fig. 2 for both the graphene sheet (main figure, for $\Omega=0.02$) and free space (inset). For this frequency, SPPs are supported by the sheet, as clearly seen in the presence of a pole in the TM components of the angular spectrum of graphene at $q_{p}\equiv k_{p}/k_{\omega} \approx 1.9$. The location of poles in complex $q-$space, $q_p$, provide the spectral region of existence of SPPs and bounded TE modes, and their corresponding wavevector, $\mathrm{Re}(q_{p})$, and propagation length, $L_{p}=\lambda / ( 2 \pi \mathrm{Im}(q_{p})) $, which are rendered in Fig. 1b.

In principle, $\hat{G}_{ij}(r)$ must be computed numerically \cite{nota_integral}. However, some heuristic properties can be extracted from the form of Eq.\ref{e3} and with the visual help of Fig. 2. Equation 1 shows that the EM field originates from the interference of a continuum of oscillatory functions, $J_{ij}^\tau(qr)$ (which are given by the symmetry of the problem), each of them launched with an amplitude  $D_{ij}^\tau$ (which depends on the material properties). For large distances, the oscillatory functions oscillate very quickly with $q$, and the contributions to the integral of ``smooth'' (in a scale $\sim 1/r$) parts of the angular spectrum cancel out. Thus, only the sharpest spectral features in $D_{ij}^\tau$ are relevant at long distances from the source. These can be divided in three classes: (i) square-root singularities, which appear for some components at the light-cone ($q=1$) and lead to a $1/r$ dependence of the field at large distances,
(ii) poles associated to bound EM modes which decay exponentially with distance both in the normal and in-plane directions and (iii) kinks (points where the derivative of the angular spectrum is discontinuous) which lead to $1/r^2$ decays.  The expressions for these contributions are given in the Supplementary Material. As we will show later, all these spectral features show up in the EM field in different spatial regions.

In what follows we present results for the electric field at the graphene sheet radiated by a nanoemitter as a function of frequency $\omega$. Let us analyze separately the emission by considering the different components of the dipole.

\begin{figure}[thb!]
\includegraphics[width=10cm]{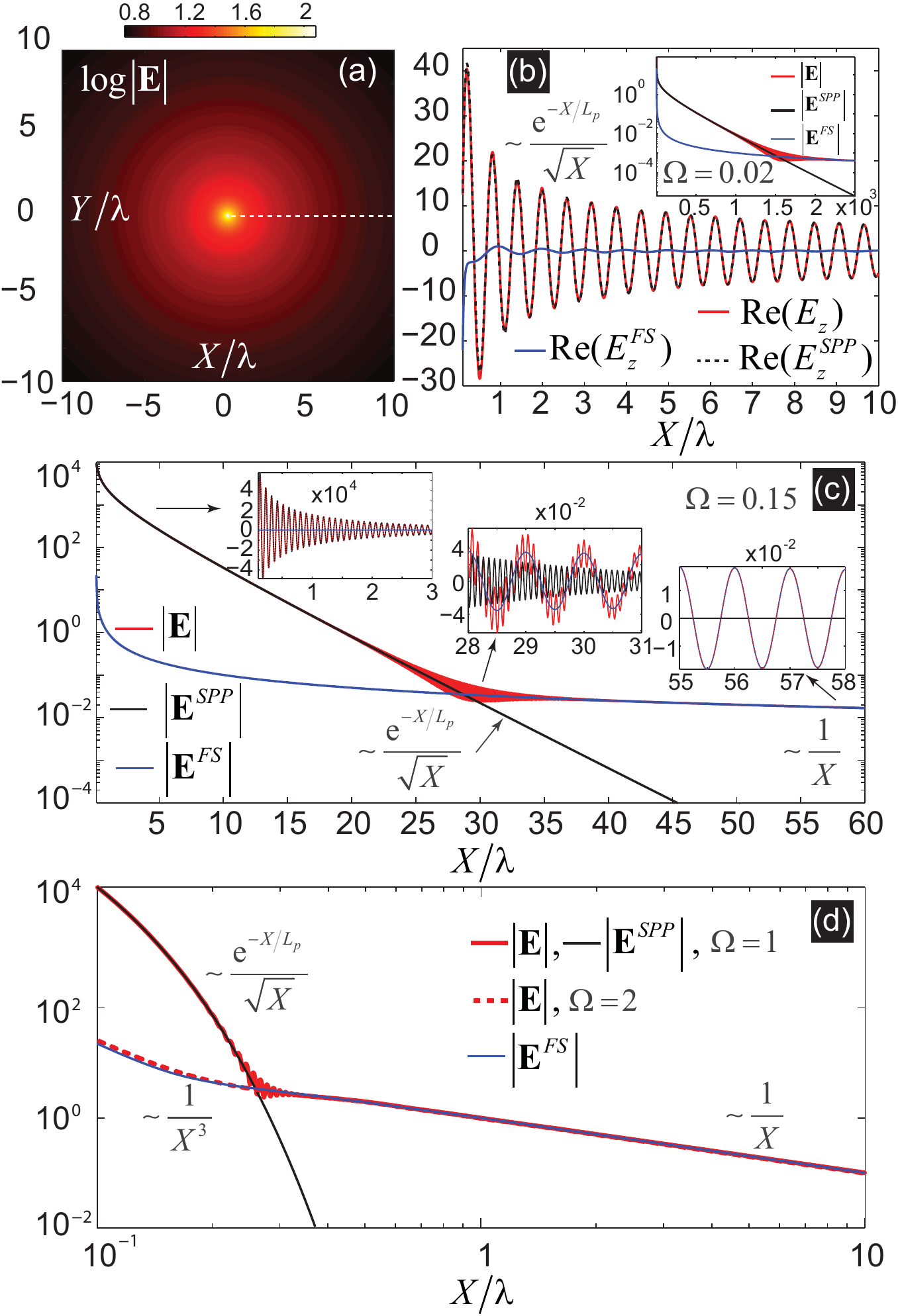}\\
\caption{Spatial dependencies of the electric field radiated by a dipole perpendicular to the graphene sheet: $\mathbf{p}=p\mathbf{e}_z$. In (a) the field colormap for $\log(|\mathbf{E}|)$ is shown for $\Omega=0.02$. The white discontinuous line represents the section along which all the 1D field dependencies are shown. In (b), (c), (d) and their insets, both $\log(|\mathbf{E}|)$ and $\mathrm{Re}(E_z)$ are plotted as function of distances for different $\Omega$. In all the panels together with the full field, both surface plasmon and free-space contributions are shown. All fields are normalized to $k_\omega p /2$, which is the largest value of the field radiated by a dipole moment in free space at distance $r=\lambda$.}\label{Figvertical}
\end{figure}

{\it Perpendicular dipole, $\mathbf{p}=p \, \mathbf{e}_z$.-}
Figure 3 presents the spatial dependence of the electric field at the graphene sheet. Due to the symmetry of the problem the field does not depend on the polar angle $\theta$, as illustrated in panel (a).
Panels (b)-(d) render the results for four different frequencies, each one being representative of a different regime. In each case, the exact electric field ($E$, red lines)
is plotted together with the SPP contribution ($E^{SPP}$, black lines) and, for reference,
the electric field that the dipole would radiate in free space (FS) ($E^{FS}$, blue lines).

Panel (b) renders the results for $\Omega=0.02$, that corresponds to a frequency much smaller than the chemical potential. In this case, $k_p \approx k_\omega $ and $L_p \gg \lambda$. As the figure shows, the exact field virtually coincides with the SPP contribution even at distances much smaller than the wavelength. At larger distances the exponentially-decaying SPP dominates until it is superseded by a algebraically-decaying field, which turns out to be well approximated by the free-space radiation (see inset to Fig 2b). The crossover distance $R_c$ typically occurs at $ R_c \sim 10 L_p$ which, for this frequency regime, may be hundreds or even thousands of wavelengths. The field oscillates with a similar period before and after $R_c$, presenting an interference pattern at distances close to $R_c$, where SPP and FS fields are of the same order.

Figure 3c is representative of an intermediate frequency range ($\Omega = 0.15$), in which $k_p \gg k_\omega $ and $L_p$ is a few times the wavelength. In this range the field can also be well approximated by the sum of two terms: the SPP (which dominates at short distances) and a "free-space" wave (prevailing at long distances). However now the spatial period of the EM field at $R<R_c$ and $R>R_c$ are vastly different (see insets to Fig. 3c). What is important to note is that now the E-field amplitude in the close proximity of the emitter is much larger (around one thousand times) than that obtained for free-space radiation. For higher frequencies, $L_p<\lambda$ and the SPP is an overdamped wave ($0.4<\Omega<1.6$ for the temperature and chemical potential considered in this paper, see Fig. 1). Still the field at the graphene sheet can be represented by a SPP at $R<R_c$ and as free-space radiation for larger distances (see Fig. 2d for $\Omega=1$). However, in this case, $R_c $ is in the subwavelength regime, and in fact it vanishes as $\Omega \rightarrow \Omega_0$ ($\Omega_{0}=1.6$ for the material parameters considered in this paper). Nevertheless, note that the huge amplification of the E-field in the vicinity of the emitter is still maintained thanks to the SPP excitation. For frequencies $\Omega>\Omega_0$, the graphene layer does not support SPPs but it does support TE surface waves. However, as represented in Fig. 2d for $\Omega=2$, the fields at the graphene sheet are very approximately the same as if the graphene layer were not present. The $TE$ field is so weakly bounded that it virtually does not couple to the point emitter.

An important message that can be extracted from Fig. 3 is that in all cases where SPPs are supported by the graphene sheet, the intensity of the field at distances from the source smaller than $R_c$ is orders or magnitude larger than in free-space. Our results show that this enhancement can be as large as $10^5-10^6$. This amplification, combined with the tunability of the chemical potential (and thus of $\Omega$) through doping levels and/or external gates may have important practical applications. For instance, whether a subwavelength region close to the emitter (and the extent of this region) is strongly illuminated or not can be externally controlled by DC voltages. For this, the chemical potential of the graphene sheet must be changed so that for the resulting $\Omega = \hbar \omega/\mu$ SPPs are either strongly overdamped or not supported at all (see Fig. 2d).

\begin{figure}[thb!]
\includegraphics[width=10cm]{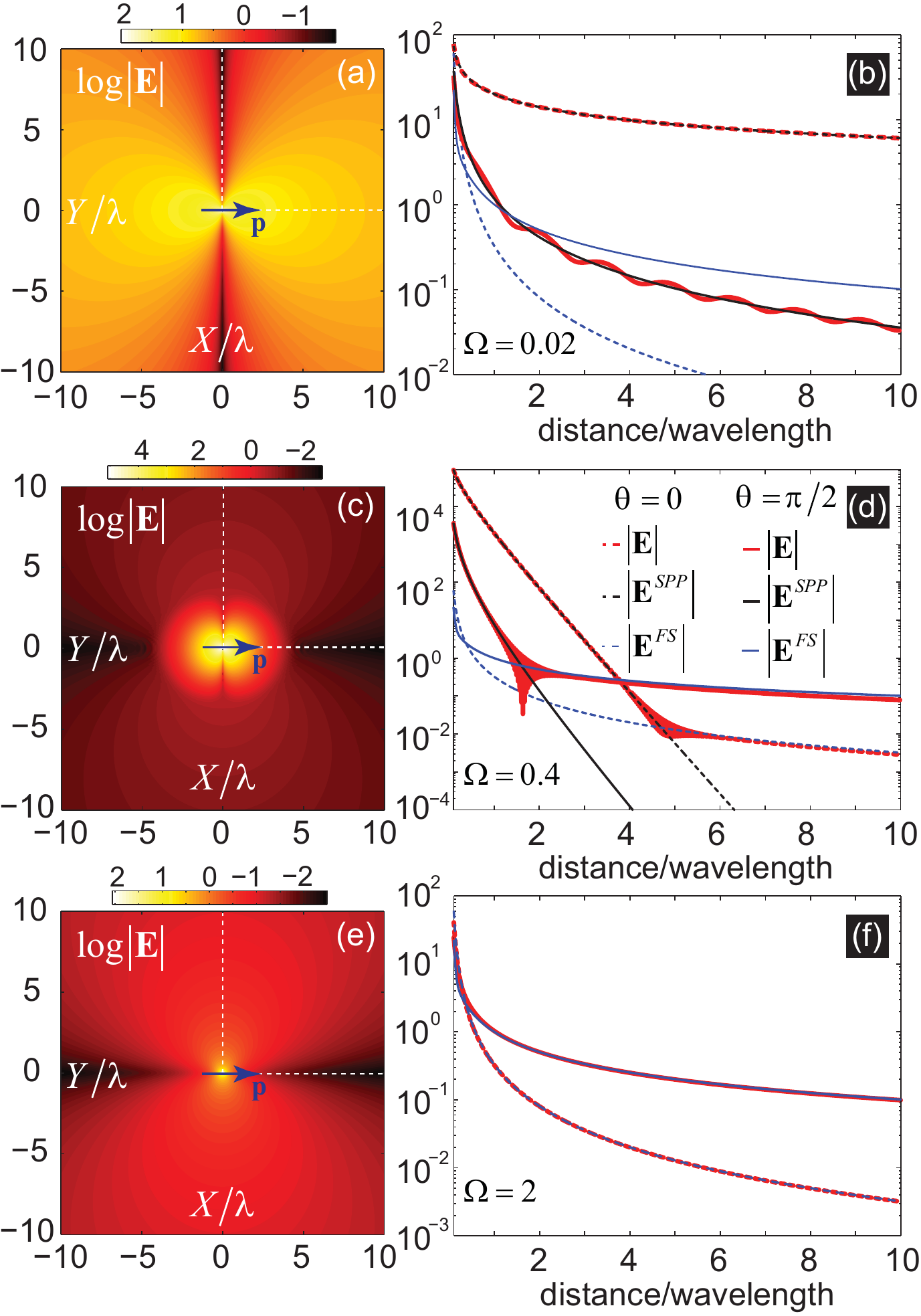}\\
\caption{The spatial dependencies of the electric field radiated by a dipole parallel to the graphene sheet $\mathbf{p}=p\mathbf{e}_x$ for different $\Omega$: $\Omega=0.02$ ($\nu=1$THz, $\lambda=0.3$mm), $\Omega=0.4$ ($\nu=19.4$THz, $\lambda=15.5\mu$m), $\Omega=2$  ($\nu=97$THz, $\lambda=3.1\mu$m). In (a), (c), (e) the field colormaps for $\log(|\mathbf{E}|)$ are shown. The white discontinuous lines represent the directions along which all the 1D field dependencies are shown in (b), (d), (f). In all the panels together with the full field, both surface plasmon and free-space contributions are shown. The normalization of the fields is the same as in Fig. 3. }\label{Fighorizontal}
\end{figure}

{\it Parallel dipole, $\mathbf{p}=p \, \mathbf{e}_x$.-}.
Let us consider now a dipole parallel to the graphene sheet, in a direction that we take at the $x-$direction (i.e. $\theta=0$). The electric field at the graphene sheet is
$\mathbf{E}(r,\theta)=(G_{rr}(r)\cos\theta,G_{\theta\theta}(r)\sin\theta,G_{zr}(r)\cos\theta)^T\,p $, where $^{T}$ stands for transposition.

In contrast to the case of the perpendicular dipole, the electric field radiated by the parallel dipole is strongly angular-dependent, and this dependency drastically changes with $\Omega$, see colormaps in Fig. 4. Overall, the field pattern follows the rules previously discussed for the vertical dipole: (i) For $\Omega<\Omega_0$, SPPs are sustained and dominate the field in an inner core region, with a size related to the SPP decay length and, thus, strongly frequency dependent. Outside this core region, the field at the graphene sheet is very approximately equal to the one in free-space (ii) For $\Omega>\Omega_0$, the core region has vanished and the field is practically the same as for free-space. However, in the case of the in-plane dipole there are two main differences. The first one is that, strictly speaking, the long-distance limit of the in-plane field is not the transversal free-space field. Instead, the asymptotic field arises from the kink in the angular spectrum  (see Fig. 2), being reminiscent of the Norton wave found in the studies of radio transmission at the earth surface \cite{Norton36} and radiation by holes in metals \cite{NikitinPRL10}.  Our calculations show that the Norton wave in graphene is predominantly TM polarized. Figure 5 renders a "phase-diagram" showing the main character of the field as a function of both frequency and distance to the source for the two directions of the point dipole considered in this paper. Notice however that the small value of the field in the region where NW dominate most probably prevents its experimental detection. The second difference is related to the polarization of the inner core and outer region. As SPPs have a longitudinal field component (thus vanishing at $\theta=\pi/2$, see Fig. 4a) and FS radiation is transversal (vanishing at $\theta=0$, see Fig 4e), the polarization patterns of the inner core and the outer region are rotated by $\pi/2$, as illustrated in Fig. 4c. Correspondingly, for $\Omega<\Omega_0$, the inner core region extends further away
 at $\theta=0$ than at $\theta=\pi/2$ (see Fig. 4d).

\begin{figure}[thb!]
\includegraphics[width=10cm]{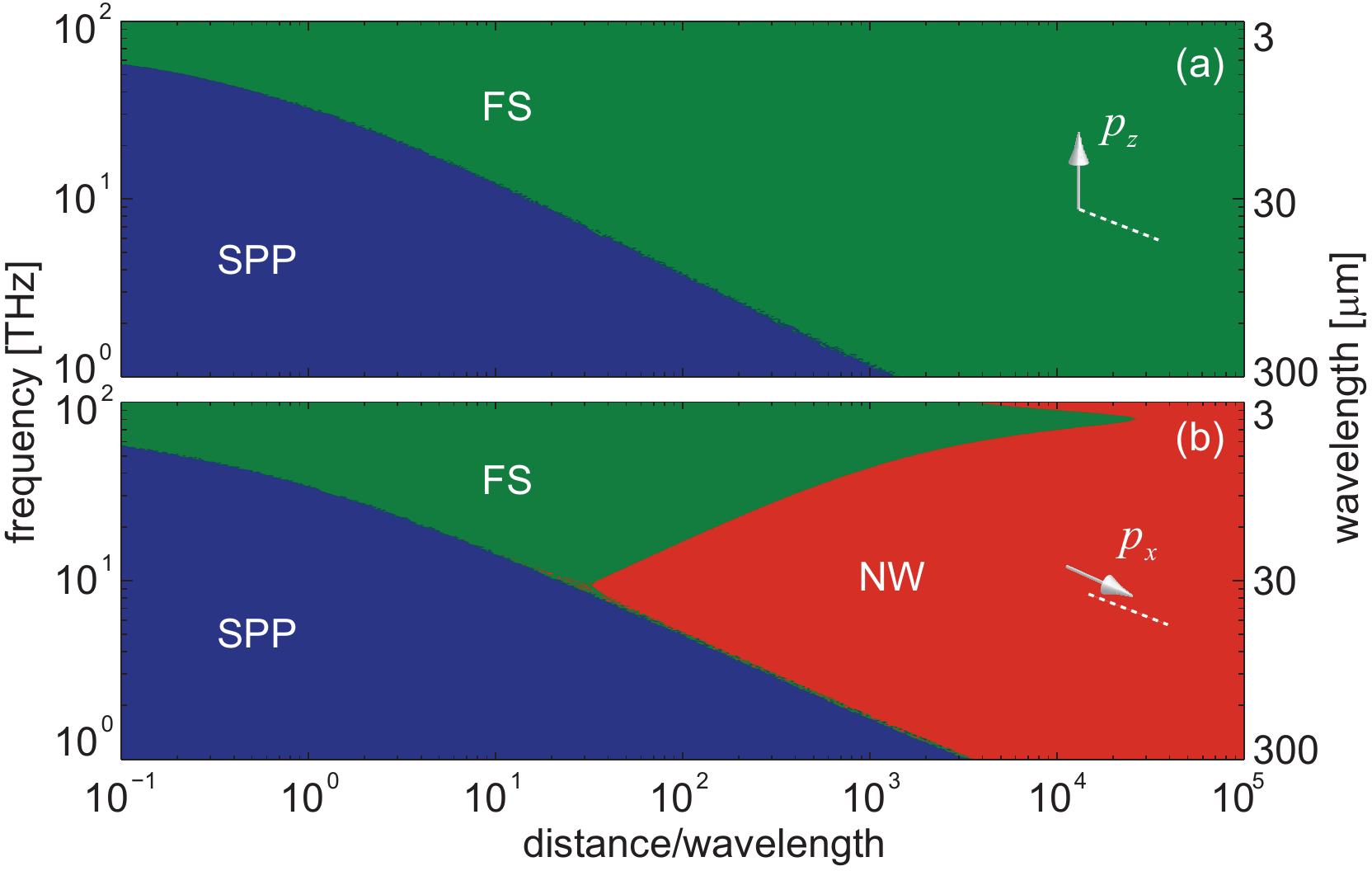}\\
\caption{Phase-diagram showing the main character of the field at the graphene sheet as a function of source frequency and distance to the source. Panel (a) is for a dipole perpendicular to the sheet, while panel (b) is for a parallel one.  In both cases T=300K and $\mu=0.2$eV.} \label{Figphase}
\end{figure}

\emph{Note added in proof:} during the process of submission of this manuscript we have become aware of the related work \emph{F. H. L. Koppens et al., arXiv:1104.2068v1 (2011)}, where the Purcell factor of the point emitter close to graphene structures has been studied.

\newpage

\appendix*

\section{\emph{Supplementary Material: Explicit expressions for the conductivity and the Green's Dyadic of the graphene sheet}}

\section{The model for the conductivity of graphene}

The conductivity of graphene has been computed within the random phase approximation \cite{Wunsch06,Hwang07,Falkovsky08}. In terms of the chemical potential $\mu$ and the temperature $T$ the conductivity can be expressed as:
\begin{equation}
\sigma = \sigma_{intra} + \sigma_{inter}
\end{equation}
where, for relaxation times much larger that $\omega^{-1}$, the intraband and interband contributions are:
\begin{equation}\label{s2}
\begin{split}
&\sigma_{intra} = \frac{2ie^2t}{\hbar\pi\Omega}\ln\left[2\cosh\left(\frac{1}{2t}\right)\right], \\
&\sigma_{inter} = \frac{e^2}{4\hbar}\left[\frac{1}{2} +\frac{1}{\pi}\arctan\left(\frac{\Omega-2}{2t}\right) - \frac{i}{2\pi}\ln\frac{(\Omega+2)^2}{(\Omega-2)^2+(2t)^2}\right].
\end{split}
\end{equation}
In this expressions $\Omega = \hbar\omega/\mu$ and $t = T/\mu $, with T expressed in units of energy.

A finite relaxation time, $\tau$, can customarily taken into account by substituting $\omega$ by $\omega+\imath \tau^{-1}$ in $\sigma_{intra}$.

\section{The electromagnetic Green's Dyadic for a 2D electron gas}

\begin{figure}[thb!]
\includegraphics[width=6cm]{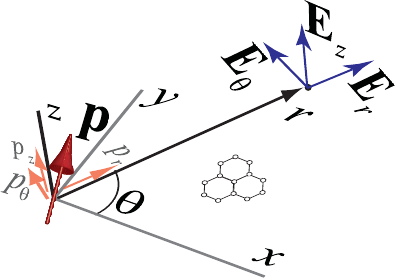}\\
\caption{The geometry of the problem: the dipole with the dipole moment $\mathbf{p}$ placed at the origin creates an electric field $\mathbf{E}$ at the point $(r,\theta,0)$. The plane of 2DEG coincides with $z=0$. Both cartesian and cylindrical coordinate systems are shown.}
\label{geom}
\end{figure}

Throughout this Supplementary Material, we express coordinates in the in-plane ($\mathbf{R}$) and normal  ($Z$) directions to the  graphene sheet in dimensionless units as $\mathbf{r}=k_\omega \mathbf{R} = 2 \pi \mathbf{R}/ \lambda$ and $z=k_\omega Z$.

The electric field $\mathbf{E}(\mathbf{r},z)$ emitted by an electric dipole, with dipole moment $\mathbf{p}(z')$ and placed at the point $(\mathbf{r}'=0,z')$, is given through the Green's dyadic (GD) $\hat{G}(\mathbf{r},z,z')=\hat{G}(\mathbf{r},z;\mathbf{r}'=0,z')$ by the following relation
\begin{equation}\label{s3}
\begin{split}
\mathbf{E}(\mathbf{r},z) = \hat{G}(\mathbf{r},z;z') \, \mathbf{p}(z').
\end{split}
\end{equation}

\subsection{General form of the Green's Dyadic under the presence of a 2D electron gas}

The GD can be expressed (see for instance \cite{Novotny,FelsenMarcuvitz}) in terms of the electromagnetic modes in free-space $\mathbf{u}_{\mathbf{q}\tau}e^{i\mathbf{qr}+iq_zz}$, characterized by their  in-plane momentum $\mathbf{q}=\mathbf{k}/k_\omega$ and polarization $\tau=\mathrm{TE, TM}$):
\begin{equation}\label{s4.0}
\begin{split}
\hat{G}(\mathbf{r},z;z')= \hat{G}_0(\mathbf{r},z;z') + \hat{G}_R(\mathbf{r},z;z'),
\end{split}
\end{equation}
with $\hat{G}_0$ being the GD in free space,
\begin{equation}\label{s4.1}
\begin{split}
\hat{G}_0(\mathbf{r},z;z')=\sum_{\tau}\int \frac{d\mathbf{q}}{2q_z}\mathbf{u}^{\pm}_{\mathbf{q}\tau}\mathbf{u}_{\mathbf{q}\tau}^{\pm T}e^{i\mathbf{\mathbf{q}\mathbf{r}}+iq_z|z-z'|},
\end{split}
\end{equation}
and $\hat{G}_R$ being the contribution due to the reflection from our 2D system
\begin{equation}\label{s4.2}
\begin{split}
\hat{G}_R(\mathbf{r},z;z')=\sum_{\tau}\int \frac{d\mathbf{q}}{2q_z}r^\tau_{q}\mathbf{u}^{+}_{\mathbf{q}\tau}\mathbf{u}_{\mathbf{q}\tau}^{- T}e^{i\mathbf{\mathbf{q}\mathbf{r}}+iq_z(z+z')}.
\end{split}
\end{equation}
In these expressions ``$+$'' corresponds to the regions where $z>z'$ while ``$-$'' is for $z<z'$, and the unitary vectors characterizing the polarization of each mode are:
\begin{equation}\label{s4.3}
\begin{split}
\mathbf{u}_{\mathbf{q}TE}^{\pm} = \frac{1}{q}
\begin{pmatrix}
-q_y\\
q_x\\
0
\end{pmatrix}, \q
\mathbf{u}_{\mathbf{q}TM}^{\pm} = \frac{q_z}{q}
\begin{pmatrix}
q_x\\
q_y\\
\mp\frac{q^2}{q_z}
\end{pmatrix}.
\end{split}
\end{equation}

In the above expressions $q_z=\sqrt{1-q^2}$ is the normalized z-component of the wavevectors and $r^\tau_{q}$ is the reflection coefficient for the 2D electron gas.
These coefficients are obtained so that the magnetic $\mathbf{H}$ and electric $\mathbf{E}$ fields satisfy the following continuity relations:
\begin{equation}\label{s13.0}
\begin{split}
&\mathbf{e}_z\times (\mathbf{E}_{-}-\mathbf{E}_{+}) = 0,\\
&\mathbf{e}_z\times(\mathbf{H}_{-}-\mathbf{H}_{+}) = \frac{4\pi}{c}\mathbf{j} = \frac{4\pi}{c}\sigma\,\mathbf{e}_z\times\mathbf{E_{+}},
\end{split}
\end{equation}
where $\mathbf{E}_{-}$ ($\mathbf{H}_{-}$) and $\mathbf{E}_{+}$ ($\mathbf{H}_{+}$) stay for the electric (magnetic) fields in the regions of negative and positive $z$, respectively, and $\mathbf{e}_z$ is the unitary vector along the $+z$ direction. As a result of the matching we have
\begin{equation}\label{s13}
\begin{split}
r^{TE}_{q} = \frac{-\alpha}{\alpha + q_{z}}, \q
r^{TM}_{q} = \frac{-\alpha q_z}{\alpha q_{z} + 1 },
\end{split}
\end{equation}
with $\alpha = 2\pi \sigma/c$, being the dimensionless 2D conductivity.

Explicitly, we have for $\hat{G}_0 = \hat{G}_0^{TM}+\hat{G}_0^{TE}$
\begin{equation}\label{s14}
\begin{split}
&\hat{G}_0^{TE}(\mathbf{r}) =\frac{ik_\omega}{8\pi^2}
\int\dfrac{d\mathbf{q}}{q_{z}q^2}e^{i\mathbf{q}\mathbf{r}+iq_z|z-z'|}
\begin{pmatrix}
q_y^2 & -q_xq_y & 0\\
-q_xq_y & q_x^2 & 0\\
0 & 0 & 0\\
\end{pmatrix},\\
&\hat{G}_0^{TM}(\mathbf{r}) =\frac{ik_\omega}{8\pi^2}
\int\dfrac{d\mathbf{q}}{q^2}e^{i\mathbf{q}\mathbf{r}+iq_z|z-z'|}
\begin{pmatrix}
q_x^2q_{z} & q_xq_yq_{z} & \mp q_xq^2\\
q_xq_yq_{z} & q_y^2q_{z} & \mp q_yq^2\\
\mp q_xq^2 & \mp q_yq^2 & q^4/q_{z}\\
\end{pmatrix}.
\end{split}
\end{equation}
Analogously, $\hat{G}_R = \hat{G}_R^{TM}+\hat{G}_R^{TE}$ reads
\begin{equation}\label{s14.1}
\begin{split}
&\hat{G}_R^{TE}(\mathbf{r}) =\frac{ik_\omega}{8\pi^2}
\int\dfrac{d\mathbf{q}}{q_{z}q^2}r^{TE}_{q}e^{i\mathbf{q}\mathbf{r}+iq_z(z+z')}
\begin{pmatrix}
q_y^2 & -q_xq_y & 0\\
-q_xq_y & q_x^2 & 0\\
0 & 0 & 0\\
\end{pmatrix},\\
&\hat{G}_R^{TM}(\mathbf{r}) =\frac{ik_\omega}{8\pi^2}
\int\dfrac{d\mathbf{q}}{q^2}r^{TM}_{q}e^{i\mathbf{q}\mathbf{r}+iq_z(z+z')}
\begin{pmatrix}
q_x^2q_{z} & q_xq_yq_{z} & q_xq^2\\
q_xq_yq_{z} & q_y^2q_{z} & q_yq^2\\
- q_xq^2 & - q_yq^2 & -q^4/q_{z}\\
\end{pmatrix}.
\end{split}
\end{equation}

\subsection{Green's Dyadic for the dipole placed at the interface $z=0$}

For the case when the dipole is placed onto the graphene sheet, that is when we take $z'=0^+$ in the previous expressions, the Green's dyadic $\hat{G}(\mathbf{r},z)=\hat{G}(\mathbf{r},z,z'=0^+)$ can be greatly simplified.

The symmetry of the problem makes it convenient to work in cylindrical coordinates $(r,\theta,z)$, see Fig.~\ref{geom}:
\begin{equation}\label{s8}
\begin{split}
x = r \cos\theta, \q y = r \sin\theta, \q z=z.
  \end{split}
\end{equation}
In this system of coordinates the Green's dyadic can be obtained from the one in cartesian coordinates through
\begin{equation}\label{s9}
\begin{split}
\hat{G}^{cylindrical} = \hat{T}^{-1}\,\hat{G}^{cartesian}\,\hat{T},
\end{split}
\end{equation}
where
\begin{equation}\label{s10}
\begin{split}
\hat{T} =
\begin{pmatrix}
\cos\theta & -\sin\theta & 0\\
\sin\theta & \cos\theta & 0\\
0 & 0 & 1\\
\end{pmatrix}.
  \end{split}
\end{equation}
In the polar coordinates the Green's Dyadic can be expressed in the form of Sommerfeld integrals involving the Angular Spectrum Dyadic $\hat{D}(q)$:
\begin{equation}\label{e3}
G_{ij}= \frac{ik_\omega}{8\pi} \sum_{\tau=TE,TM} \int_0^\infty \, dq  \,
D_{ij}^\tau(q) \, J_{ij}^\tau(qr)e^{iq_zz},
\end{equation}
with
\begin{equation}\label{e4}
\begin{split}
 &\hat{J}^{TE}(q r) =\begin{pmatrix}
J_+(qr)& 0 & 0\\
0 & J_-(qr) & 0\\
0 & 0 & 0\\
\end{pmatrix}, \q \hat{J}^{TM}(q r) =
\begin{pmatrix}
J_-(qr) & 0 & J_1(qr)\\
0 & J_+(qr) & 0\\
J_1(qr) & 0 & J_0(qr)\\
\end{pmatrix}.
\end{split}
\end{equation}

In this expressions, $J_\pm(qr)= J_0(qr)\pm J_2(qr)$ and $J_n(qr)$ is the Bessel function of $n$th order. The non-zero components of the angular spectrum dyadic are
$D^{TE}_{rr}= D^{TE}_{\theta \theta}=t^{TE}_{q}q/q_{z}$, $D^{TM}_{rr}= D^{TM}_{\theta \theta}=t^{TM}_{q}qq_{z}$, $D^{TM}_{r z}= D^{TM}_{z r}=-2 i q^2 t^{TM}_{q}/q_{z}$, and $D^{TM}_{zz}= 2 q^{3} t^{TM}_q/q_{z}$.

\subsection{SPP contribution to the field}

The contribution from the SPP is calculated from the residues of the poles in the TM part of the Green's dyadic. For the case of a vertical dipole, this calculation has been performed in Ref.~\onlinecite{Hansonw08}.
In order to find the SPP field, all Bessel functions must be expressed in terms of Hankel functions of first ($H^{(1)}$) and second ($H^{(2)}$) kind. Then the contour integral is deformed and closed either in the upper half-plane of complex variable $q$ (for terms involving $H^{(1)}$) or in the lower half-plane (terms where $H^{(2)}$ appears).
As the SPP pole is enclosed only when closing the integration contour in the half-plane, the final result is
\begin{equation}\label{s16}
\begin{split}
 &\hat{G}^{SPP}(r,z) =\frac{k_\omega}{8}
\frac{q_p}{\alpha^3} e^{iq_{pz}z}
\begin{pmatrix}
H^{(1)}_-(q_pr) & 0 & 2iq_p\alpha H^{(1)}_1(q_pr)\\
0 & H^{(1)}_+(q_pr) & 0\\
2iq_p\alpha H^{(1)}_1(q_pr) & 0 & 2q_p^2\alpha^2 H^{(1)}_0(q_pr)\\
\end{pmatrix},
\end{split}
\end{equation}
where $H^{(1)}_\pm(qr)= H^{(1)}_0(qr)\pm H^{(1)}_2(qr)$ and $H^{(1)}_n(qr)$ being the first-kind Hankel function of $n$th order and $q_p = \sqrt{1-\alpha^{-2}}$ with $q_{pz}=-\alpha^{-1}$ are the normalized k-vector components of the SPP.

\subsection{Asymptotic of the Green's dyadic at the surface $z=0$: the leading order}

For large values of $r$, such that $r\gg1$ the asymptotic expansion of GD can be performed.
As in the previous subsection, the limits of the integral \eqref{e3} are extended to the whole real $q-$axis using the Hankel functions, and then the long-distance asymptotic value of these functions is taken. The result reads
\begin{equation}\label{s17}
\begin{split}
&\hat{G}(r,z) = \frac{k_\omega e^{i\frac{\pi}{4}}}{8\pi}\sqrt{\frac{2}{\pi r}}\int\limits_{-\infty}^\infty dq \sqrt{q}e^{iqr+iq_{z}z}
\begin{pmatrix}
\frac{q_z}{\alpha q_{z} + 1}& 0 & -\frac{q}{\alpha q_{z} + 1}\\
0 & \frac{1}{\alpha + q_{z}} & 0\\
-\frac{q}{\alpha q_{z} + 1} & 0 & \frac{q^2}{q_z(\alpha q_{z} + 1)}\\
\end{pmatrix}.
\end{split}
\end{equation}

On the plane $z=0$,  the main contribution is expected to proceed from the branch point $q_z=0$. Then the computational procedure consists in transforming the integral to such a form so that the integrand is proportional to $1/q_z$. The element $zz$ already contains this factor, so that the transformation for it is not necessary.  The other elements should be integrated by parts. After the transformation the integrand is expanded in the vicinity of $q_z=0$ and then using
\begin{equation}\label{s18}
\begin{split}
\int\limits_{-\infty}^\infty dq \frac{e^{iqr}}{q_z} = \pi H^{(1)}_0(qr),
\end{split}
\end{equation}
 the integration is trivially performed:
\begin{equation}\label{s19}
\begin{split}
\hat{G}(r,0) = \frac{k_\omega e^{i\frac{\pi}{4}}}{8}\sqrt{\frac{2}{\pi r}}
\begin{pmatrix}
\frac{1}{ir}& 0 & \frac{\alpha}{ir}\\
0 & \frac{i}{r\alpha^2} & 0\\
\frac{\alpha}{ir} & 0 & 1\\
\end{pmatrix}H^{(1)}_0(qr).
\end{split}
\end{equation}
Finally, substituting $H^{(1)}_0(qr)$ by its asymptotic value, we arrive at
\begin{equation}\label{s20}
\begin{split}
\hat{G}(r,0) = \frac{k_\omega }{4\pi r}
\begin{pmatrix}
\frac{1}{ir}& 0 & \frac{\alpha}{ir}\\
0 & \frac{i}{r\alpha^2} & 0\\
\frac{\alpha}{ir} & 0 & 1\\
\end{pmatrix}e^{ir}.
\end{split}
\end{equation}
The dependency $\sim 1/r$ is exactly the same as for the case of free space (spherical-type wave in vacuum) while the $\sim 1/r^2$ decay is redolent of the Norton wave arising in the radio transmission at the earth surface \cite{Norton36} and radiation by holes in metals \cite{NikitinPRL10}.

\subsection{Analytical form of the Green's Dyadic of free space for $z=0$}

The free space GD follows from Eq.~\eqref{s4.0} just by setting $\alpha=0$ (in which case, the reflection coefficient vanishes) so that we simply have $\hat{G}=\hat{G}_0$.
$\hat{G}_0$ can be rearranged as
\begin{equation}\label{s5}
\begin{split}
\hat{G}_0(\mathbf{r},z) = \frac{ik_\omega}{8\pi^2}\int \frac{d\mathbf{q}}{q_z}
\begin{pmatrix}
1-q_x^2 & -q_xq_y & - q_xq_z\\
-q_xq_y & 1-q_y^2 & - q_yq_z\\
- q_xq_z & - q_yq_z & 1-q_z^2\\
\end{pmatrix}e^{i\mathbf{\mathbf{q}\mathbf{r}}+iq_zz}.
  \end{split}
\end{equation}
Using the Green's function for a scalar potential
\begin{equation}\label{s6}
\begin{split}
G_0(\rho) = \frac{k_\omega}{4\pi}\frac{e^{i\rho}}{\rho}, \q \rho=r^2+z^2.
  \end{split}
\end{equation}
Eq.~\eqref{s5} can be written in a differential form
\begin{equation}\label{s7}
\begin{split}
\hat{G}_0(\mathbf{r},z) =
\begin{pmatrix}
1+\partial_{x}^2 & \partial_{x}\partial_{y} & \partial_{x}\partial_{z}\\
\partial_{x}\partial_{y} & 1+\partial_{y}^2 & \partial_{y}\partial_{z}\\
\partial_{x}\partial_{z} & \partial_{y}\partial_{z} & 1+\partial_{z}^2\\
\end{pmatrix}G(\rho).
  \end{split}
\end{equation}
After performing the derivatives in \eqref{s7}, the result in cylindrical coordinates is:
\begin{equation}\label{s11}
\begin{split}
&\hat{G}_0^{cylindrical}(r,0) = \frac{k_\omega}{4\pi}\cdot\frac{e^{ir}}{r^3}
\begin{pmatrix}
2(-1+ir) & 0 & 0\\
0 & -1+ir + r^2 & 0\\
0 & 0 & -1+ir + r^2\\
\end{pmatrix}.
  \end{split}
  \end{equation}


\begin{thebibliography}{99}

\bibitem{Novoselov05}
K. S. Novoselov, A. K. Geim, S. V. Morozov, D. Jiang, M. I., Katsnelson,
I. V. Grigorieva, S. V. Dubonos, and A. A. Firsov, Nature \textbf{438}, 197 (2005).

\bibitem{Review09}
A. H. Castro Neto, F. Guinea, N. M. R. Peres, K. S. Novoselov and A. K. Geim, Rev. Mod. Phys. \textbf{81}, 109 (2009).

\bibitem{Geim09}
A. K. Geim, Science, \textbf{324}, 1530 (2009).

\bibitem{Bonaccorso10}
F. Bonaccorso, Z. Sun, T. Hasan and A. C. Ferrari, Nature Photonics, \textbf{4}, 611 (2010).

\bibitem{Casiraghi07}
C. Casiraghi, A. Hartschuh, E. Lidorikis, H. Qian, H. Harutyunyan, T. Gokus, K. S. Novoselov and A. C. Ferrari,
Nano Lett. \textbf{7}, 2711(2007).

\bibitem{Blake07}
P. Blake, E. W. Hill, A. H. Castro Neto, K. S. Novoselov, D. Jiang, R. Yang,
T. J. Booth and A. K. Geim, Appl. Phys. Lett. \textbf{91}, 063124 (2007).

\bibitem{Shung86}
K. W. -K. Shung, Phys. Rev. B \textbf{34}, 979 (1986).

\bibitem{Campagnoli89}
G. Campagnoli and E. Tosatti, in Progress on Electron Properties of metals, edited by R. Girlanda et al., Kluwer, Dordrecht, p. 337 (1989).

\bibitem{Vafek06}
O. Vafek, Phys. Rev. Lett. \textbf{97}, 266406 (2006).

\bibitem{Hansonw08}
G. W. Hanson, J. Appl. Phys. \textbf{103}, 064302 (2008).

\bibitem{MikhalkovPRL07}
S. A. Mikhalkov and K. Ziegler, Phys. Rev. Lett. \textbf{99}, 016803 (2007).

\bibitem{Bludov10}
Yu. V. Bludov, M. I. Vasilevskiy and N. M. R. Peres, Europhys. Lett. \textbf{92}, 68001 (2010).

\bibitem{Engheta11}
A. Vakill and N. Engheta, arXiv:1101.3585v1 (2011).

\bibitem{alejandro11}
A. Gonzalez-Tudela, D. Martin-Cano, E. Moreno, L. Martin-Moreno, C. Tejedor and F.J. Garcia-Vidal, Phys. Rev. Lett. {\bf 106}, 020501 (2011).

\bibitem{Wunsch06}
B. Wunsch, T. Stauber, F. Sols and F. Guinea, New J. of Phys. \textbf{8}, 318 (2006).

\bibitem{Hwang07}
E. H. Hwang and S. Das Sarma, Phys. Rev. B \textbf{75}, 205418 (2007).

\bibitem{Falkovsky08}
L. A. Falkovsky, Physics-Uspekhi \textbf{51}, 887 (2008).

\bibitem{NikitinPRL10}
A. Yu. Nikitin, F.J. Garcia-Vidal, and L. Martin-Moreno, Phys. Rev. Lett. \textbf{105}, 073902 (2010).

\bibitem{FelsenMarcuvitz}
L. P. Felsen and N. Marcuvitz, \textit{Radiation and Scattering of Waves}
(IEEE Press, Piscataway, NJ, 1994).

\bibitem{nota_integral} Although this is a notoriously difficult task, due to the  presence of strongly oscillatory
factors and singularities (branch points and poles)
in the integrand, there are special but standard techniques to compute them, see \cite{FelsenMarcuvitz}.

\bibitem{Novotny}
L. Novotny and B. Hetch, \textit{Principles of Nano-Optics} (Cambridge University Press, New York, 2006).

\bibitem{Norton36}
K. A. Norton, \textit{Proc. IRE}, 24, 1367, (1936).



\end{thebibliography}
\end{document}